\documentclass[preprint, 5p]{elsarticle}

\usepackage{amsmath, amsthm, amssymb}
\usepackage{amsfonts}
\usepackage{graphicx}
\usepackage{mathtools}
\usepackage[version=3]{mhchem}
\usepackage{booktabs}
\usepackage{multirow}

\newcommand{\NAME}{\mathrm{diag}}
\newcommand{\PNAME}{\mathrm{hybrid}}
\renewcommand{\O}{\mathcal{O}}
\newcommand{\HXC}{\mathrm{HXC}}
\newcommand{\XC}{\mathrm{XC}}
\newcommand{\loc}{\mathrm{loc}}
\newcommand{\LDOS}{\mathrm{LDOS}}
\newcommand{\R}{\mathbb{R}}
\newcommand{\rr}{\mathbf{r}}
\DeclareMathOperator{\Ker}{Ker}
\DeclareMathOperator{\Ran}{Ran}
\DeclareMathOperator{\Span}{Span}

\DeclarePairedDelimiterX\braket[2]{\langle}{\rangle}{#1\,\delimsize\vert\,\mathopen{}#2}

\begin{document}

\begin{frontmatter}
\title{Preconditioning Magnetic Systems in Kohn-Sham Density Functional Theory}

\author{Clémentine Barat\textsuperscript{1, 2, 3}}
\author{Antoine Levitt\textsuperscript{3}}
\author{Marc Torrent\textsuperscript{1, 2}}

\address{\textsuperscript{1}CEA, DAM, DIF, F-91297 Arpajon, France}
\address{\textsuperscript{2}Université Paris-Saclay, CEA, Laboratoire Matière en Conditions Extrêmes, F-91680 Bruyères-le-Châtel, France}
\address{\textsuperscript{3}Laboratoire de Mathématiques d'Orsay, Université Paris-Saclay, 307 rue Michel Magat, 91400 Orsay, France}

\begin{abstract}
	The convergence of the self-consistent field iterations in Kohn-Sham density functional theory can be significantly hindered by the presence of small eigenvalues in the dielectric matrix, which are often associated with electronic phase transitions in magnetic systems. In this work, we study this type of convergence issues and propose a new preconditioning scheme to mitigate them. Our preconditioning scheme is inspired by the Stoner model and based on a non-interacting susceptibility that neglects orbital variations. We demonstrate the effectiveness of our approach on a range of ferromagnetic systems, showing that it can significantly reduce the number of iterations required to achieve convergence in the vicinity of magnetic phase transitions.
\end{abstract}

\begin{keyword}
Preconditioning \sep Kohn-Sham DFT \sep Self-consistent field \sep Magnetism \sep Convergence
\end{keyword}
\end{frontmatter}

\section{Introduction}

Kohn-Sham density functional theory (KS-DFT) is a widely used method in solid-state physics and chemistry for simulating the electronic properties of materials~\cite{hohenberg_inhomogeneous_1964, kohn_self-consistent_1965}. The Kohn-Sham equations are typically solved using self-consistent field (SCF) iterations, and improving the convergence of these iterations has been an ongoing challenge since the first numerical resolution of these equations.

The ground-state density in Kohn-Sham DFT is the solution of a nonlinear equation $R(\rho) = 0$, where $R$ is the fixed-point residual. The differential of $R$ is given by the adjoint dielectric matrix $\varepsilon^\dagger = 1 - \chi_0 K_\HXC$, with $\chi_0$ denoting the non-interacting susceptibility and $K_\HXC$ the Hartree-exchange-correlation kernel. Methods to accelerate the convergence of SCF iterations generally involve constructing approximations of this differential, to apply a quasi-Newton method. Two distinct approaches exist to build these approximations, and in practice, both are combined.

The first approach consists of generic methods that do not rely on prior knowledge of $R$. Instead, they use the iteration history to construct approximate differentials. In the context of SCF iterations in Kohn-Sham DFT, these methods are referred to as \textit{mixing}. Some widely used mixing schemes include Broyden's method~\cite{broyden_class_1965, srivastava_broydens_1984} and Anderson acceleration~\cite{anderson_iterative_1965}, the latter being equivalent to Pulay's DIIS mixing~\cite{pulay_convergence_1980}. All of these fall under the same category of secant or multisecant methods. The development and refinement of these algorithms have been an active area of research, with numerous variants and improvements proposed in the literature (e.g.,~\cite{johnson_modified_1988, kresse_efficient_1996, eyert_comparative_1996, marks_robust_2008, marks_predictive_2021, banerjee_periodic_2016, herbst_robust_2022, das_accelerating_2023, ge_acceleration_2025}).

The second approach consists of KS-DFT-specific methods that use simple physical models or knowledge of the structure of $R$ to construct estimates of the adjoint dielectric matrix. These methods are known as \textit{preconditioning} techniques. 
The earliest proposition of a preconditioner for Kohn-Sham DFT is the Kerker preconditioner~\cite{kerker_efficient_1981}, which approximates the dielectric response with the Thomas-Fermi screening function. Another widely used preconditioner is the Resta preconditioner, which uses the model dielectric function for semiconductors from~\cite{resta_thomas-fermi_1977}, also based on the Thomas-Fermi model. 

Many improvements to these two preconditioners have been proposed since.
The Kerker and Resta preconditioners, initially designed for plane-wave calculations, have been adapted for real-space implementations or other more complex frameworks~\cite{shiihara_real-space_2008,kumar_preconditioning_2020, kim_resta-like_2022}. 
Significant efforts have also been directed toward developing preconditioners suitable for inhomogeneous systems. For instance, Lin and Yang~\cite{lin_elliptic_2013} proposed to use an elliptic partial differential equation with inhomogeneous coefficients, while Zhou et al.~\cite{zhou_applicability_2018} proposed a generalization of the Kerker preconditioner for mixed systems. In~\cite{herbst_black-box_2020}, Herbst and Levitt developed a parameter-free preconditioner based on the local density of states (LDOS).
Alternative strategies include the use of the full dielectric matrix computed on reduced systems~\cite{ho_dielectric_1982, anglade_preconditioning_2008}. 
Beyond linear preconditioners, nonlinear approaches also exist. These involve solving a simplified problem at each SCF iteration, such as the Thomas-Fermi-von Weizsäcker equation for inhomogeneous systems~\cite{raczkowski_thomas-fermi_2001}, a second-variational prediction operator~\cite{sawamura_second-variational_2004}, or an auxiliary bosonic system~\cite{hasnip_auxiliary_2015}.

Dederichs and Zeller~\cite{dederichs_self-consistency_1983} and later Herbst and Levitt~\cite{herbst_black-box_2020} identified three sources of slow convergence related to the dielectric operator of the system: charge sloshing in large simulation cells caused by the long range of the Coulomb interaction, localized orbitals that yield large contributions in the non-interacting susceptibility, and lastly, phase transitions associated with a non-invertible dielectric matrix, typically in magnetic systems. Most of the available preconditioners are designed to mitigate charge sloshing issues. In this work, we focus on the convergence issue associated with magnetic phase transitions. We propose a hybrid preconditioner $P^\PNAME = 1 - \chi_0^\LDOS K_H - \chi_0^\NAME K_\XC$, which extends the LDOS preconditioner of~\cite{herbst_black-box_2020} by incorporating an additional term $\chi_0^\NAME$ in which we neglect the orbital variations. This approach is inspired by the Stoner model, to address convergence challenges near ferromagnetic phase transitions.

This paper is organized as follows: Section~\ref{sec:SCF} reviews Kohn-Sham DFT, specifically its magnetic formulation in the collinear-spin DFT formalism, and discusses the convergence properties of SCF iterations. Section~\ref{sec:convergence} details the specific convergence issues related to magnetism, including a more general discussion on the impact of small outlier eigenvalues in linear systems. Section~\ref{sec:preconditioner} introduces our proposition for a preconditioning scheme, inspired by the Stoner model, to reduce convergence issues specific to magnetic systems. Lastly, Section~\ref{sec:results} presents numerical results for our preconditioner.

\section{The self-consistent field iterations in spin-pola\-rized Kohn-Sham DFT}\label{sec:SCF}

Consider a system of $N$ electrons in a domain $\Omega$ evolving in a potential $V_\mathrm{ext}$ created by the nuclei and the core electrons. In the Kohn-Sham density functional theory (KS-DFT) framework, these electrons are treated as non-interacting particles in a density-dependent effective potential $V_\HXC = V_H + V_\XC$. 
To allow for spin-polarization, we work within spin-DFT, where the exchange-correlation potential $V_\XC$ depends not only on the total electron density but also on its spin-resolved components. A common restriction is the collinear spin approximation, in which the exchange-correlation potential acts independently on each spin channel. Under this assumption, the spin, $\alpha \in \{+, -\}$, becomes a good quantum number, leading to the following simplified formulation of Kohn-Sham equations for collinear-spin-DFT~\cite{barth_local_1972, jacob_spin_2012}:
\begin{align}	
\begin{split}
	& \left[ -\frac12 \nabla^2 + V_\mathrm{ext} + V_\HXC^\alpha(\rho^+, \rho^-) \right] \psi_i^\alpha = \epsilon_i^\alpha \psi_i^\alpha \\
	&\int_\Omega \overline{\psi_i^\alpha} \psi_j^\alpha = \delta_{ij} \\
	& \rho^\alpha = \sum_{i=1}^\infty f(\epsilon_i^\alpha - \epsilon_F) |\psi_i^\alpha|^2 \\
	& V_\HXC^\alpha(\rho^+, \rho^-) = v_c(\rho^++\rho^-) + V_\XC^\alpha(\rho^+, \rho^-) \\
	& \int_\Omega \rho^+ + \rho^- = N
\end{split}
	\label{eq:KS}
\end{align}
for $i \geq 1$ and $\alpha \in \{+, -\}$.
Here $f(x) = 1/(1+\exp(x/T))$ is the Fermi-Dirac function at smearing temperature
$T$, $v_c$ is the Coulomb kernel, and $V_\XC^\alpha$ is the exchange-correlation potential for spin channel $\alpha \in \{+, -\}$. We will denote $\rho = \begin{pmatrix} \rho^+\\ \rho^- \end{pmatrix}$ and $V = \begin{pmatrix} V^+\\ V^- \end{pmatrix}$ the spin-resolved densities and potentials. The finite temperature is not physical here, it serves as a regularization in periodic simulations to compensate for the k-point discretization. Note also that we do not explicitly write the k-point dependence of the Kohn-Sham orbitals.

Following the formalism used in~\cite{herbst_black-box_2020}, we define the potential-to-density mapping $F$ by
\begin{align}
	F^\alpha\left(V\right) = \sum_{i=1}^\infty f(\epsilon_i^\alpha - \epsilon_F) |\psi_i^\alpha|^2
\end{align}
where, for $\alpha \in \{+, -\}$, $(\epsilon_i^\alpha, \psi_i^\alpha)$ are the normalized eigenpairs of $ -\frac12 \nabla^2 + V_\mathrm{ext} + V^\alpha $ and the Fermi level $\epsilon_F$ is adjusted so that $ \sum_{\alpha \in \{+, -\}} \sum_{i=1}^\infty f(\epsilon_i^\alpha - \epsilon_F) = N$. This allows us to rewrite the Kohn-Sham equations~\eqref{eq:KS} as the following fixed-point, or self-consistent, equation:
\begin{align}
	F(V_\HXC(\rho)) = \rho .
	\label{eq:KSFP}
\end{align}

This equation is often solved using self-consistent field (SCF) iterations. One of the simplest types of SCF iterations is damped iterations, where the density is updated with $\rho_{n+1} = \rho_n + \tau [ F(V_\HXC(\rho_n)) - \rho_n ]$ for some damping parameter $0<\tau \leq 1$.
As is standard (see~\cite{herbst_black-box_2020, woods_computing_2019}), the linearization of these iterations, around a solution $\rho_*$, with an optimal $\tau$ yields a convergence rate $\theta = (\kappa_s(\varepsilon^\dagger)-1)/(\kappa_s(\varepsilon^\dagger)+1)$ where $\kappa_s(\varepsilon^\dagger) = \lambda_{\max}(\varepsilon^\dagger)/\lambda_{\min}(\varepsilon^\dagger)$ is the spectral condition number, i.e., the ratio of the largest to the smallest eigenvalue, of the adjoint dielectric matrix
\begin{align}
	\varepsilon^\dagger = 1 - \chi_0 K_\HXC .
\end{align}
Here, $\chi_0 = D_V F(V_\HXC(\rho_*))$ is the non-interacting susceptibility, and $K_\HXC = D_\rho V_\HXC(\rho_*)$ denotes the Hartree-exchange-correlation kernel. Because $\chi_0$ is self-adjoint and non-positive, $\varepsilon^\dagger$ has a real spectrum and its spectral condition number is well-defined (see~\cite{herbst_black-box_2020}).

Within the collinear-spin formalism, these derivatives can be expressed as $2 \times 2$ Jacobian matrices of operators:
\begin{align}
	\chi_0 = 
	\begin{pmatrix} 
		\chi_0^{++} & \chi_0^{+-} 
		\\ 	\chi_0^{-+} & \chi_0^{--}
	\end{pmatrix}
\end{align}
where the matrix elements are derived within perturbation theory~\cite{adler_quantum_1962, wiser_dielectric_1963, herbst_black-box_2020, levitt_screening_2020}. 

The Hartree-exchange-correlation kernel can be decomposed into two contributions, $K_\HXC = K_H + K_\XC$, with
\begin{align}
	K_H = 
	\begin{pmatrix} 
		v_c & v_c 
		\\ 	v_c & v_c 
	\end{pmatrix} .
\end{align}

\section{Convergence issues in magnetic systems}\label{sec:convergence}

The adjoint dielectric matrix $\varepsilon^\dagger = 1 - \chi_0 (K_H + K_\XC)$ is ill-conditioned when it has either large or small eigenvalues. Since the non-interacting susceptibility $\chi_0$ is non-positive and the Hartree kernel $K_H$ is positive semi-definite, the operator $1-\chi_0 K_H$ has all its eigenvalues larger than or equal to 1. Hence, convergence issues associated with small eigenvalues of $\varepsilon^\dagger$ originate from strong exchange-correlation effects~\cite{herbst_black-box_2020}. This is typically the case in magnetic systems, especially near phase transitions.

Indeed, when a non-zero magnetization appears continuously as a function of one of the system parameters, $\gamma$, this transition must be associated with a non-invertible dielectric matrix. Due to symmetries, spin-polarized solutions of the Kohn-Sham equation are degenerate. However, if the dielectric matrix were invertible, the implicit function theorem would guarantee a locally unique solution to the Kohn-Sham equation with respect to $\gamma$, thereby preventing the emergence of multiple distinct solutions at this point. 

In this section, we first discuss quantitatively how a few small eigenvalues can affect the convergence of SCF iterations, especially with advanced mixing strategies. Then, we study magnetic phase transitions, first with the Stoner model and then in Kohn-Sham DFT.

\subsection{Convergence issues associated with small eigenvalues}\label{sec:convergence_gmres}

When the Kohn-Sham fixed-point equation is solved with optimally damped iterations, the convergence rate is given by $\theta = (\kappa_s-1)/(\kappa_s+1)$. For large values of $\kappa_s$, this rate can be approximated by $\theta \approx 1-2/\kappa_s $ and the number of iterations required to reach a tolerance $e$ is $n_\mathrm{ite} \approx (\kappa_s/2)\log(1/e)$. This implies that if the smallest eigenvalue of $\varepsilon^\dagger$ is reduced by a given factor, the number of iterations increases by this same factor. Thus, small outlier eigenvalues of $\varepsilon^\dagger$ severely degrade the convergence. However, damped iterations are rarely used in practice, as more sophisticated and more efficient fixed-point solvers are available. One such method is Anderson acceleration, whose convergence properties and limitations are discussed in~\cite{chupin_convergence_2021}.

Anderson acceleration, when applied to a linear problem with an untruncated history, is equivalent to the Generalized Minimal Residual (GMRES) method~\cite{saad_gmres_1986}, a Krylov subspace technique (see, e.g.,~\cite{walker_anderson_2011}). This section therefore focuses on the impact of small eigenvalues on the convergence of GMRES, for a matrix $A = \varepsilon^\dagger$.

Krylov subspace methods~\cite{saad_iterative_2003, driscoll_potential_1998} construct iterative approximate solutions $x_n$ to the linear problem $Ax=b$, within the subspace $x_0 + \mathcal{K}_n(A, r_0)$, where $\mathcal{K}_n(A, r_0)$ is the order-$n$ Krylov subspace generated by $A$ and the initial residual $r_0$. Equivalently, Krylov methods can be viewed as constructing approximations whose residuals can be expressed as
\begin{align}
	r_n = P_n(A) r_0
\end{align}
where $P_n$ is a polynomial of degree at most $n$ that satisfies~$P_n(0)=1$. We call the space of such polynomials $\mathcal{P}_n$.

Within this framework, the GMRES method finds the polynomial in $\mathcal{P}_n$ that minimizes the $L_2$-norm of the residual~\cite{saad_iterative_2003}, that is:
\begin{align}
		\|r_n\| = \min_{P \in \mathcal{P}_n} \|P(A) r_0\| .
		\label{eq:gmres_min}
\end{align}
In the linearization of the SCF procedure, the matrix $A$ is the adjoint dielectric matrix $\varepsilon^\dagger$, which is diagonalizable with a non-negative real spectrum. This gives the following bound: 
\begin{align}
	\|r_n\| \leq \kappa(V) \|r_0\| \min_{P \in \mathcal{P}_n} \max_{\lambda \in \mathrm{spec}(A)} |P(\lambda)| 
\end{align}
where $\kappa(V) = \|V\|\|V^{-1}\|$ is the condition number of $V$, the eigenvector matrix of $A$.

Chebyshev polynomials provide a classical bound for the above minimization problem when the spectrum of $A$ is contained in an interval $[\lambda_{\min} , \lambda_{\max}]$ with $0 < \lambda_{\min} $~\cite{saad_iterative_2003}. In this case, the minimum
\begin{align}
	\min_{P \in \mathcal{P}_n} \max_{\lambda \in [\lambda_{\min}, \lambda_{\max}]} |P(\lambda)| 
\end{align}
is achieved by a scaled and shifted Chebyshev polynomial of degree $n$. It yields the following bound for the residual:
\begin{align}
	\|r_n\| \leq 2 \kappa(V) \|r_0\| \left( \frac{\sqrt{\kappa_s} - 1}{\sqrt{\kappa_s} + 1} \right)^n \label{eq:cvg_cheby}
\end{align}
with $\kappa_s = \lambda_{\max}/\lambda_{\min}$ being the spectral condition number of $A$. The derivation of this upper bound from Chebyshev polynomials is detailed in~\cite[Section~6.11]{saad_iterative_2003}. This bound is relevant when the spectrum of $A$ is uniformly distributed over the interval $[\lambda_{\min}, \lambda_{\max}]$. However, if $A$ has outlier eigenvalues, this bound becomes overly pessimistic, as polynomials do not need to attenuate the entire interval $[\lambda_{\min}, \lambda_{\max}]$.

The impact of spectral outliers on the convergence of Krylov methods is nicely explained in~\cite{driscoll_potential_1998} from a polynomial approximation perspective. 

\begin{figure}
		\includegraphics[width=\linewidth]{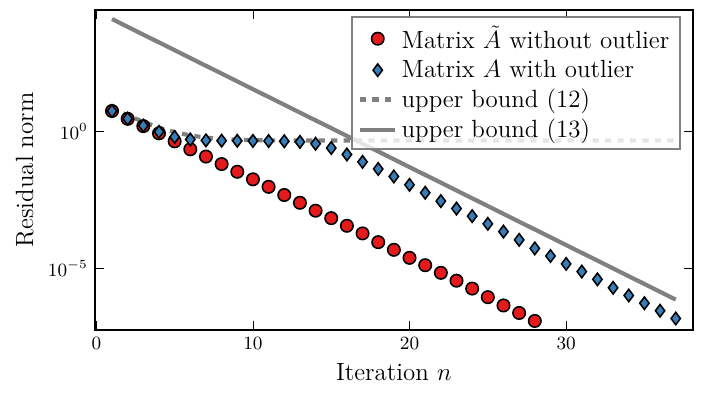}
		\caption{Residual norm of the GMRES method applied to a matrix $\tilde{A}$ with a spectrum contained in $[1, 10]$ and to a matrix $A$ with one eigenvalue equal to 0.01 and the remaining spectrum in $[1, 10]$, with the two upper bounds~\eqref{eq:gmres_bound1} and~\eqref{eq:gmres_bound2} for the latter.}
    \label{fig:gmres_outlier}
\end{figure}

For simplicity, consider a positive definite symmetric matrix $A$ whose real spectrum includes a small outlier eigenvalue $\lambda_\mathrm{out}$: \newline
\setlength{\unitlength}{\linewidth}
\begin{picture}(1, 0.2)(0, -0.1)
	\put(0, 0){\line(1, 0){\linewidth}}

	\put(0.1, -0.025){\line(0, 1){0.05}}
	\put(0.09, -0.07){$0$}

	\put(0.13, -0.025){\line(0, 1){0.05}}
	\put(0.12, 0.05){$\lambda_\mathrm{out}$}

	\put(0.5, -0.025){\line(0, 1){0.05}}
	\put(0.49, 0.05){$\lambda_2$}

	\put(0.51, -0.025){\line(0, 1){0.05}}
	\put(0.52, -0.025){\line(0, 1){0.05}}
	\put(0.56, -0.025){\line(0, 1){0.05}}
	\put(0.59, -0.025){\line(0, 1){0.05}}
	\put(0.61, -0.025){\line(0, 1){0.05}}
	\put(0.62, -0.025){\line(0, 1){0.05}}
	\put(0.63, -0.025){\line(0, 1){0.05}}
	\put(0.67, -0.025){\line(0, 1){0.05}}
	\put(0.7, -0.025){\line(0, 1){0.05}}
	\put(0.71, -0.025){\line(0, 1){0.05}}
	\put(0.74, -0.025){\line(0, 1){0.05}}
	\put(0.78, -0.025){\line(0, 1){0.05}}
	\put(0.8, -0.025){\line(0, 1){0.05}}
	\put(0.82, -0.025){\line(0, 1){0.05}}
	\put(0.83, -0.025){\line(0, 1){0.05}}
	\put(0.86, -0.025){\line(0, 1){0.05}}

	\put(0.7, 0.05){$\cdots$}

	\put(0.9, -0.025){\line(0, 1){0.05}}
	\put(0.89, 0.05){$\lambda_{\max}$}
\end{picture}
We denote $\tilde{A}$ the restriction of $A$ to the eigensubspace associated with the spectrum of $A$ excluding $\lambda_\mathrm{out}$. Suppose the GMRES method is applied to solve a linear system associated with $\tilde{A}$, yielding a sequence of polynomials $(P_n)$ and residuals $\tilde{r}_n = P_n(\tilde{A}) \tilde{r}_0$ that achieve a convergence rate $\theta$. We consider here that it means $\|P_n(\tilde{A}) \tilde{r}_0 \| \approx \theta^n \|\tilde{r}_0\|$. Now consider the two polynomials $P_n$ and $Q_n(X) = (1-X/\lambda_\mathrm{out})P_{n-1}(X)$, both of which belong to $\mathcal{P}_n$. Since the GMRES method minimizes the residual norm over $\mathcal{P}_n$ (equation~\eqref{eq:gmres_min}), these polynomials yield the two upper bounds $\|r_n\| \leq \|P_n(A) r_0\|$ and $\|r_n\| \leq \|Q_n(A) r_0\|$, where $r_n$ is the residual of the GMRES method applied to $A$. Assuming the initial residual reads $r_0 = s_0 + \tilde{r}_0$, where $s_0$ is an eigenvector associated with $\lambda_\mathrm{out}$, we have
\begin{align}
	&\|r_n\|  \leq |P_n(\lambda_\mathrm{out})| \|s_0\| + \| P_n(A) \tilde{r}_0\| \approx \|s_0\| + \theta^{n} \|\tilde{r}_0\| \label{eq:gmres_bound1} \\
	&\|r_n\| \leq \|Q_n(A) \tilde{r}_0\| \lesssim \frac{\lambda_{\max}}{\lambda_\mathrm{out}} \theta^{n-1} \|\tilde{r}_0\| \label{eq:gmres_bound2} ,
\end{align}
where we have used $P_n(\lambda_\mathrm{out}) \approx 1$, following from the condition $P_n(0)=1$, and $Q_n(A) r_0 = Q_n(A) \tilde{r}_0$, following from $Q_n(\lambda_\mathrm{out}) = 0$.
These two bounds are illustrated in Figure~\ref{fig:gmres_outlier}. They suggest that the convergence curves for this type of system initially follow the convergence rate of the matrix without outliers, until the residual along the mode associated with $\lambda_\mathrm{out}$ dominates, and then exhibit a plateau of $n_\mathrm{plateau}$ iterations, before aligning again with the convergence of the matrix without outliers. Here $n_\mathrm{plateau}$ is the number of additional iterations needed for the residual on the remaining spectrum in the second bound (equation~\eqref{eq:gmres_bound2}) to match the residual on the remaining spectrum in the first bound (equation~\eqref{eq:gmres_bound1}), that is
\begin{align}
	\theta^{n_\mathrm{plateau}-1} \approx \frac{\lambda_\mathrm{out}}{\lambda_{\max}} .
		\label{eq:n_plateau}
\end{align}
This means that the number of additional steps introduced by the small outlier $\lambda_\mathrm{out}$ corresponds to the number of iterations needed to reduce the residual by a factor $\frac{\lambda_{\max}}{\lambda_\mathrm{out}}$ for the matrix $\tilde{A}$. 

This argument can be readily extended to the case of a matrix with multiple outliers, where the length of the plateau will correspond to the sum of the above estimates for each (effective) outlier. 

Thus, while Anderson acceleration reduces the impact of small outlier eigenvalues of the dielectric matrix, it does not eliminate it entirely. The number of required additional iterations scales logarithmically with the outlier, with a convergence that is all the more affected when outliers are numerous and when the underlying matrix (without outliers) is ill-conditioned. This analysis also provides a diagnostic for convergence issues associated with a plateau, linking them to the presence of small eigenvalues in the adjoint dielectric matrix.

\subsection{Stoner model}

The Stoner model provides a simple framework to understand convergence issues associated with magnetic transitions, as discussed in~\cite{dederichs_self-consistency_1983}.

The Stoner model for ferromagnetism~\cite{stoner_collective_1938, zeller_spin-polarized_2006} gives a scalar description for the total magnetization $M$ of a system. In this model, we consider that the total magnetization $M$ rigidly shifts a prescribed density of states by $IM$ between the $+$ and $-$ spin channels. The Stoner parameter $I$ characterizes the strength of the exchange interaction in the system. This leads to the following fixed-point equation
for $M$:
\begin{align}
	G(M) = M
\label{eq:stoner}
\end{align}
with
\begin{align}
	\begin{split}
		G(M) = & \int_{-\infty}^{\epsilon_F(M)} D\left(\epsilon + \frac12 IM\right) d\epsilon \\
		& - \int_{-\infty}^{\epsilon_F(M)} D\left(\epsilon - \frac12 IM\right) d\epsilon ,
	\end{split}
\end{align}
where $D$ is the density of states (per spin channel) of the non-spin-polarized system, and $\epsilon_F(M)$ is determined by the constraint on the total number of electrons, $\int_{-\infty}^{\epsilon_F(M)} D(\epsilon + IM/2) \,d\epsilon + \int_{-\infty}^{\epsilon_F(M)} D(\epsilon - IM/2) \,d\epsilon = N$.

Equation~\eqref{eq:stoner} has a trivial solution $M_0=0$. Since $M$ is bounded by the total number of electrons, a sufficient condition for a magnetic solution to exist is $G'(0) > 1$. This condition rewrites $ID(\epsilon_F^0) > 1$, with $\epsilon_F^0=\epsilon_F(0)$, and is the well-known Stoner criterion.

The fixed-point equation~\eqref{eq:stoner} can be interpreted as the stationary condition for the following Stoner energy:
\begin{align}
	\begin{split}
		E_\mathrm{Stoner}(M) = &\int_{-\infty}^{\epsilon_F(M)+\lambda(M)} \epsilon D\left(\epsilon + \frac12 IM\right) d\epsilon \\
		&+ \int_{-\infty}^{\epsilon_F(M)-\lambda(M)} \epsilon D\left(\epsilon - \frac12 IM\right) d\epsilon \\
		&+ \frac14 IM^2
	\end{split}
\end{align}
in which $\epsilon_F(M)$ and $\lambda(M)$ are determined by the conditions on the total number of electrons and total magnetization:
\begin{align}
	&\begin{split}
	N = &\int_{-\infty}^{\epsilon_F(M) + \lambda(M)} D\left(\epsilon + \frac12 IM\right) d\epsilon \\ &+ \int_{-\infty}^{\epsilon_F(M) - \lambda(M)}D\left(\epsilon - \frac12 IM\right)d\epsilon \\
	\end{split} \\
	&\begin{split}
	M = &\int_{-\infty}^{\epsilon_F(M) + \lambda(M)} D\left(\epsilon + \frac12 IM\right) d\epsilon \\ &- \int_{-\infty}^{\epsilon_F(M) - \lambda(M)}D\left(\epsilon - \frac12 IM\right)d\epsilon .
	\end{split}
\end{align}
The stationary condition imposes $\lambda(M)=0$, which recovers the Stoner fixed-point equation. This energy gives a stability condition for the non-spin-polarized solution $M_0=0$, which is $E_\mathrm{Stoner}''(0) = (2D(\epsilon_F^0))^{-1}(1-ID(\epsilon_F^0)) > 0$. This is not fulfilled when the Stoner criterion is satisfied. In this case, the system will spontaneously polarize in spin.

In order to understand more precisely how the magnetic solution appears, we use a perturbation expansion of $G(M)$ around 0, which allows us to rewrite equation~\eqref{eq:stoner} as
\begin{align}
	0 = M(ID(\epsilon_F^0)-1 + a M^2) + \O(M^4)
\label{eq:stoner_lin}
\end{align}
with $a = \left(D''(\epsilon_F^0) - 3D'(\epsilon_F^0)^2/D(\epsilon_F^0) \right)I^3/24$. The sign of $a$ determines two distinct scenarios.

For $a < 0$, equation~\eqref{eq:stoner_lin} has two non-zero solutions $M_\pm = \pm (\sqrt{-a})^{-1}\sqrt{ID(\epsilon_F^0)-1} + \O({(M_\pm)}^2)$ when the Stoner criterion is met, which are actual minima of the energy, i.e., $E_{\mathrm{Stoner}}''(M_\pm) + \O({(M_\pm)}^3) > 0$. The magnetic solution therefore appears continuously with respect to $I$, so the system exhibits a second-order magnetic transition.

For $a >0$, equation~\eqref{eq:stoner_lin} has two non-zero solutions $M_\pm = \pm (\sqrt{a})^{-1}\sqrt{1 - ID(\epsilon_F^0)} + \O({(M_\pm)}^2)$ when the Stoner criterion is not met and those solutions are maxima of the Stoner energy ($E_\mathrm{Stoner}''(M_\pm) + \O({(M_\pm)}^3) < 0$). The linearization of equation~\eqref{eq:stoner} does not give a magnetic solution when the Stoner criterion is met, but in this case, the Stoner fixed-point equation has at least one magnetic solution. However, this magnetic solution does not appear continuously with respect to $I$, resulting in a first-order magnetic transition. 

\begin{figure}
	\centering
	\includegraphics[width=\linewidth]{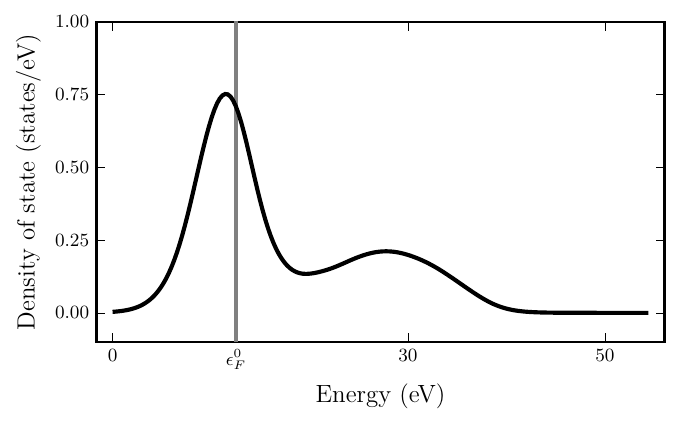}
	\includegraphics[width=\linewidth]{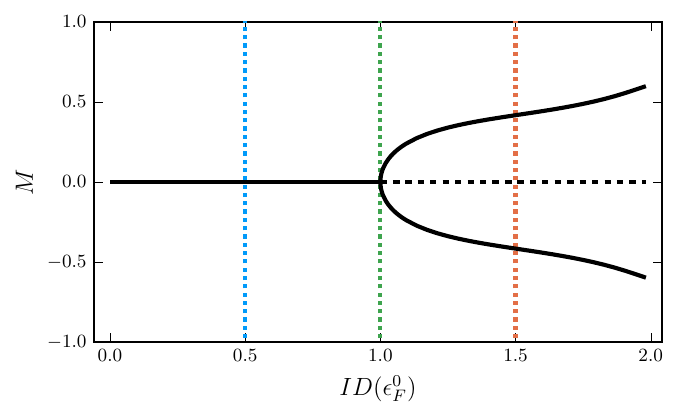}
	\includegraphics[width=\linewidth]{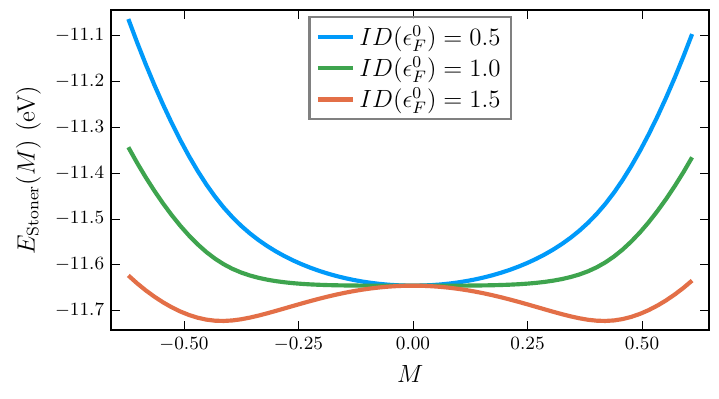}
	\caption{Stoner model in the case $a<0$. From top to bottom: Model density of states, magnetization predicted by the Stoner model with respect to the Stoner parameter, and corresponding energy surfaces for $ID(\epsilon_F^0) = 0.5$, $ID(\epsilon_F^0) = 1.0$ and $ID(\epsilon_F^0) = 1.5$.}
    \label{fig:stoner2}
\end{figure}

Figure~\ref{fig:stoner2} shows an example of the magnetization predicted by the Stoner model for a given density of states in the case $a<0$, which yields a second-order magnetic transition. We have also represented a few energy surfaces that illustrate the correspondence between the local extrema and the predicted magnetization.

The scalar fixed-point equation $M=G(M)$ can be solved using fixed-point iterations. The convergence of this method is determined by the derivative of $G$ at the fixed point (see~\cite{dederichs_self-consistency_1983}), and the analogue of the dielectric matrix in this simple model is $\varepsilon_\mathrm{Stoner}(M) = 1 - G'(M)$. At the Stoner criterion ($ID(\epsilon_F^0) = 1$), we have $\varepsilon_\mathrm{Stoner}(0)=0$; thus, for systems exhibiting a second-order magnetic transition, convergence can be arbitrarily slow near this point. For systems exhibiting a first-order magnetic transition, the transition occurs before reaching the Stoner criterion, which mitigates convergence issues.

\subsection{Magnetic transitions in Kohn-Sham DFT}

In Kohn-Sham DFT, we also expect second-order magnetic phase transitions to be associated with zero eigenvalues of the dielectric matrix.

Consider a parameter-dependent Kohn-Sham equation of the form
\begin{align}
	R(\rho, \gamma) = 0
\end{align}
where $R(\rho, \gamma) = \rho - F^\gamma(V^\gamma_\HXC(\rho))$ represents the fixed-point residual of $\rho$ at the parameter $\gamma$ (e.g. pressure, volume, electronic temperature\ldots). The differential of $R$ with respect to $\rho$ at a solution $\rho_*(\gamma)$ is the adjoint dielectric matrix $\varepsilon^\dagger(\gamma)$. As previously discussed, a magnetic solution can only appear continuously with respect to $\gamma$ at points where $\varepsilon^\dagger(\gamma)$ is not invertible. We assume that $\rho_*(0)$ is non-magnetic and that its associated adjoint dielectric matrix $\varepsilon^\dagger(0)$ has a non-degenerate, isolated, zero eigenvalue associated with the spin mode $\rho_1$.

The bifurcation of the solution at $\gamma=0$ can be analyzed using a perturbation expansion of $R$ around $(\rho_*(0), 0)$, and a Lyapunov-Schmidt reduction~\cite{nirenberg_topics_2001}. The idea is to separate the direction along which the implicit function theorem fails from the complementary directions, along which it still applies. We denote $E_1 = \Ker(\varepsilon^\dagger(0)) = \Span (\rho_1)$, $E_2 = \Ran(\varepsilon^\dagger(0))$. Since $\varepsilon^\dagger(0)$ is diagonalizable with an isolated zero eigenvalue, $E_1$ and $E_2$ form a direct-sum decomposition of the density space. We denote $\Pi_{E_1}$, $\Pi_{E_2}$ the corresponding projectors. Writing $\rho = \rho_*(0) + t\rho_1 + \rho_2$, with $t \in \R$ and $\rho_2 \in E_2$, the equation $R(\rho, \gamma) = 0$ can be decomposed into:
\begin{align}
		& \Pi_{E_1} R (\rho_*(0) + t\rho_1 + \rho_2, \gamma) = 0  \\
		& \Pi_{E_2} R (\rho_*(0) + t\rho_1 + \rho_2, \gamma) = 0 .
\end{align}
The differential of $\Pi_{E_2} R$ with respect to $\rho_2$ at $(t, \gamma, \rho_2) = (0, 0, 0)$ is invertible, as the restriction of $\varepsilon^\dagger(0)$ to $E_2$. Then, for $t$ and $\gamma$ small enough, the implicit function theorem ensures that the second equation has a unique solution, denoted here $\rho_2(t, \gamma)$. Substituting this into the first equation reduces it to an equation in $t$ and $\gamma$ alone, $f(t, \gamma)=0$, with $f(t, \gamma) = \Pi_{E_1} R(\rho_*(0) + t\rho_1 + \rho_2(t, \gamma), \gamma)$ and where $\frac{d}{dt}f(0, 0) = \Pi_{E_1} \varepsilon^\dagger(0) \rho_1 = 0$. Using the spin inversion symmetries, we can show that $f(0, \gamma) = 0$, and $\frac{d^2}{dt^2}f(0, 0) = 0$. This leads to an equation analogous to that obtained in the perturbation expansion of the Stoner model:
\begin{align}
	0 = t(a\gamma + bt^2) + \O(t^4 + \gamma t^2 + \gamma^2t)
\end{align}
with $a=\frac{d^2}{dt \, d\gamma}f(0, 0)$ and $b=\frac16 \frac{d^3}{dt^3}f(0, 0)$, both assumed to be non-zero. We recognize a \textit{pitchfork} bifurcation, where the magnetization emerges as the square root of the parameter $\gamma$, with $t = \pm \sqrt{-(a/b) \gamma} + \O(\gamma)$, provided that $-(a/b) \gamma > 0$. 

Besides, note that $\frac{d}{dt}\rho_2(0, 0) = 0$, implying that the magnetic solution reads 
\begin{align} 
	\rho_*(\gamma) = \rho_*(0) \pm \sqrt{- (a/b) \gamma} \rho_1 + \O(\gamma) .
\end{align}
This shows that, at the transition, the spin component of the density aligns with the eigenvector associated with the zero eigenvalue of the adjoint dielectric matrix.

In order to illustrate this type of transition and the associated convergence issues, we have done some numerical experiments on the ferromagnetic transition of FCC iron. The calculations were performed with the density-functional toolkit (DFTK)~\cite{herbst_dftk_2021}. We have used a plane-wave basis with an energy cutoff of 50 Hartree and a Monkhorst-Pack k-point grid of size $20\times20\times20$, with norm-conserving pseudopotentials~\cite{van_setten_pseudodojo_2018}, the Perdew-Burke-Ernzerhof (PBE)~\cite{perdew_generalized_1996} exchange-correlation functional, and the Fermi-Dirac occupation scheme with a temperature of 0.01 Hartree. To induce the magnetic transition, we varied the lattice constant around a reference primitive lattice constant for FCC iron (2.58 Å).

\begin{figure}
	\centering
	\includegraphics[width=0.66\columnwidth]{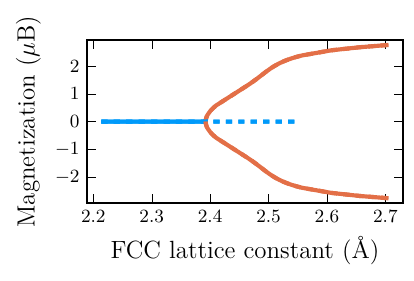}
	\includegraphics[width=0.66\columnwidth]{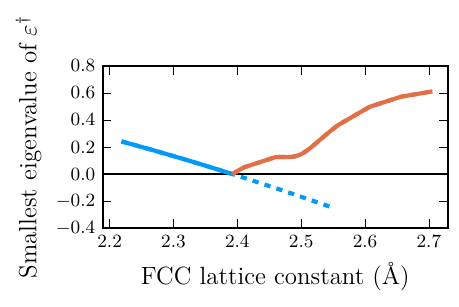}
	\includegraphics[width=0.66\columnwidth]{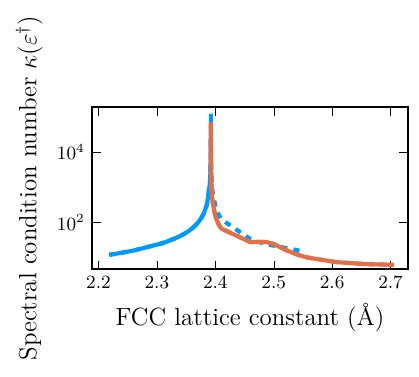}
	\caption{Total magnetization, smallest eigenvalue of the dielectric matrix, and condition number of the dielectric matrix of FCC iron, with respect to the lattice constant. The two colors represent different solution branches. The minimal energy solution is represented with a solid line.}
    \label{fig:magn_eigmin_kappa_wrt_lsf_high_temp}
\end{figure}

Figure~\ref{fig:magn_eigmin_kappa_wrt_lsf_high_temp} shows the magnetization as a function of the lattice constant for this FCC iron system, as well as the smallest eigenvalue and condition number of the adjoint dielectric matrix $\varepsilon^\dagger$. We observe that the smallest eigenvalue of the dielectric matrix decreases as the lattice constant increases, eventually reaching zero and triggering a phase transition. At this transition, the spectral condition number of the dielectric matrix blows up, which should cause convergence issues.

\begin{figure}
	\centering
	\includegraphics[width=\linewidth]{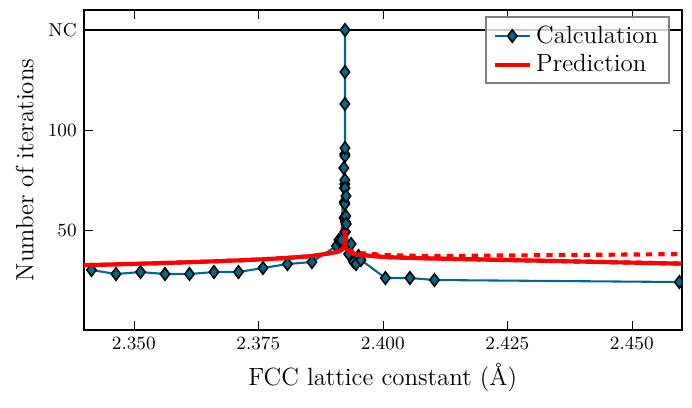}
	\caption{Number of SCF iterations needed to reach a density residual smaller than $10^{-10}$ for FCC iron, and predicted number of iterations $n_\mathrm{prediction} = 1 + \log(10^{-10}\lambda_1/\lambda_{\max}) / \log(\theta)$, with respect to the lattice constant. The calculations used Anderson acceleration with a history of length $m=10$ and no preconditioning. NC indicates that the system did not converge after 150 iterations.}
    \label{fig:nite_wrt_lsf_high_temp}
\end{figure}

\begin{figure}
	\centering
	\includegraphics[width=\linewidth]{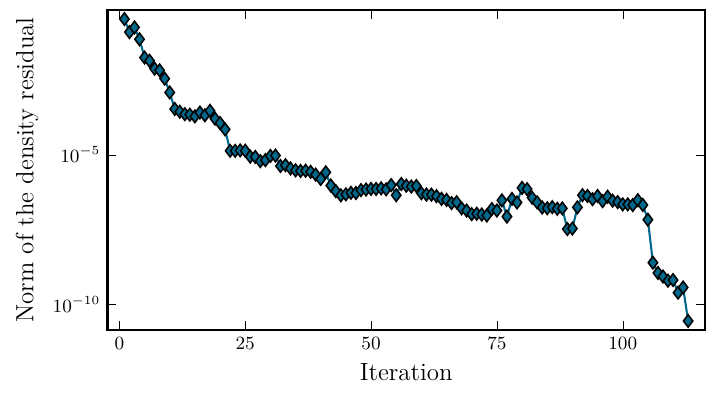}
	\caption{Residual norm of the density throughout the SCF iterations for FCC iron with a lattice constant of 2.3923913 Å. The calculations used Anderson acceleration with a history of length $m=10$ and no preconditioning.}
    \label{fig:convergence_noprecond_Fe_high_temp_transition}
\end{figure}

These convergence issues are illustrated in Figure~\ref{fig:nite_wrt_lsf_high_temp}, which shows the number of SCF iterations needed to converge the density of FCC iron for various lattice constants. An increase in the number of iterations is observed around the transition, where the adjoint dielectric matrix becomes singular. We also include the predicted number of iterations derived in Section~\ref{sec:convergence_gmres}, $n_\mathrm{prediction} = n_\mathrm{asymptotic} + n_\mathrm{plateau}$, with $n_\mathrm{plateau} = 1 + \log(\lambda_1/\lambda_{\max}) / \log(\theta)$ obtained from equation~\eqref{eq:n_plateau}, and $n_\mathrm{asymptotic} = \log(10^{-10}) / \log(\theta)$, where $\theta = (\sqrt{\lambda_{\max}/\lambda_2}-1)/(\sqrt{\lambda_{\max}/\lambda_2}+1)$ is the convergence rate that excludes the transition eigenvalue (equation~\eqref{eq:cvg_cheby}). Here, $\lambda_1 \leq \cdots \leq \lambda_{\max}$ are the eigenvalues of $\varepsilon^\dagger$. Because the number of iterations scales logarithmically with the smallest eigenvalue of $\varepsilon^\dagger$, the convergence issues remain highly localized around the transition, even though the estimation based on GMRES appears to be optimistic. 
Figure~\ref{fig:convergence_noprecond_Fe_high_temp_transition} shows an example of a convergence plot for FCC iron at the transition, on which we see the characteristic plateau caused by the small eigenvalue of the dielectric matrix.

\section{Preconditioning SCF iterations with a diagonal approximation of $\chi_0$}\label{sec:preconditioner}

Preconditioning the SCF iterations in KS-DFT consists in replacing the residuals $\rho_n - F(V_\HXC(\rho_n))$ with the preconditioned residuals
\begin{align}
	P^{-1}(\rho_n - F(V_\HXC(\rho_n)))
\end{align}
in the SCF iterations. The matrix $P$ is referred to as the preconditioner. The differential associated with this residual is now the preconditioned adjoint dielectric matrix $P^{-1}\varepsilon^\dagger$. Thus, to improve the conditioning of the SCF iterations, $P$ should approximate the adjoint dielectric matrix, $\varepsilon^\dagger = 1- \chi_0 K_\HXC$.

For semi-local exchange-correlation functionals, the $K_\HXC$ operator is relatively computationally inexpensive, whereas the operator $\chi_0$ is too expensive to use exactly in practice. In~\cite{herbst_black-box_2020}, the authors propose to approximate directly the non-interacting susceptibility $\chi_0$ instead of the inverse dielectric matrix. This shifts the focus of developing effective preconditioners to finding a computationally efficient yet relevant approximation for $\chi_0$. The well-known Kerker preconditioner~\cite{kerker_efficient_1981} falls into this category, as it approximates $\chi_0$ by a constant (cf.~\cite{herbst_black-box_2020}), as does the LDOS preconditioner, proposed in the aforementioned article, that uses an approximation based on the local density of states.

When its expression is derived using perturbation theory~\cite{herbst_black-box_2020}, the non-interacting susceptibility can be decomposed into three contributions, 
\begin{align}
	\chi_0 = \chi_0^{\delta \epsilon} + \chi_0^{\delta \epsilon_F} + \chi_0^{\delta \psi} ,
\end{align}
arising respectively from the variations of the eigenvalues, the Fermi level, and the Kohn-Sham orbitals:
\begin{align}
		&\left( \chi_0^{\delta \epsilon} \right)^{\alpha \beta}(\rr, \rr') = \delta_{\alpha \beta} \sum_{i=1}^\infty f'(\epsilon_i^\alpha-\epsilon_F) \rho_{ii}^\alpha(\rr) \rho_{ii}^\alpha(\rr') \\
		&\left( \chi_0^{\delta \epsilon_F} \right)^{\alpha \beta}(\rr, \rr') = \frac1{D^++D^-} D_\loc^\alpha(\rr) D_\loc^\beta(\rr')\\
		&\left( \chi_0^{\delta \psi} \right)^{\alpha \beta}(\rr, \rr') = \delta_{\alpha \beta} \sum_{\substack{i, j=1 \\i \neq j}}^\infty f_{ij}^\alpha \rho_{ij}^\alpha(\rr) \overline{\rho_{ij}^\alpha}(\rr') ,
\end{align}
with 
\begin{align}
	f_{ij}^\alpha = \begin{cases}
		\dfrac{f(\epsilon_i^\alpha - \epsilon_F) - f(\epsilon_j^\alpha - \epsilon_F)}{\epsilon_i^\alpha - \epsilon_j^\alpha} & \text{if } \epsilon_i^\alpha \neq \epsilon_j^\alpha, \\[1ex]
		f'(\epsilon_i^\alpha - \epsilon_F) & \text{if } \epsilon_i^\alpha = \epsilon_j^\alpha,
	\end{cases}
\end{align}
for $\alpha$, $\beta$ in $\{+, -\}$.  In these expressions, we use the orbital codensities $\rho_{ij}^\alpha(\rr) = \psi_i^\alpha(\rr)\overline{\psi_j^\alpha}(\rr)$, the density of states $D^\alpha = - \sum_i f'(\epsilon_i^\alpha - \epsilon_F)$, and the local density of states $D_\loc^\alpha(\rr) = -\sum_i f'(\epsilon_i^\alpha - \epsilon_F) \rho_{ii}^\alpha(\rr)$. 

Among these three terms, the most computationally intensive contribution when applying $\chi_0$ to a vector is by far $\chi_0^{\delta \psi}$.

In this section, we suggest a preconditioner aimed at addressing convergence issues associated with ferromagnetism. Drawing inspiration from the Stoner model in which only the eigenvalues are affected by magnetization, we propose to neglect the computationally demanding contribution $\chi_0^{\delta \psi}$.
Therefore, we propose the following model for $\chi_0$:
\begin{align}
	\chi_0^\NAME = \chi_0^{\delta \epsilon} + \chi_0^{\delta \epsilon_F} ,
\end{align}
which keeps only the diagonal contributions of Kohn-Sham orbital interactions. 

This approximation generalizes the Stoner model. Consider a non-magnetic, self-consistent, KS-DFT reference density $\rho^\pm_*  = \sum_{i=1}^\infty f(\epsilon_i - \epsilon_F) \rho_{ii}$ and the associated model non-interacting susceptibility $\chi_0^\NAME$.
We study the following linear fixed-point equation for a variation of the density $\delta \rho$ with respect to the reference density:
\begin{align}
	\begin{split}
		&\delta V = K_\mathrm{Stoner} \delta \rho \\
		&\delta \rho = \chi_0^\NAME \delta V ,
	\end{split}
\end{align}
where the potential-to-density map is given by $\chi_0^\NAME$ and the density-to-potential map is given by the Stoner model, with $\left( K_\mathrm{Stoner} \delta \rho \right)^\alpha = -\alpha I M/2$ and the magnetization $M = \int_\Omega \delta \rho^+ - \delta \rho^-$.
This leads to:
\begin{align}
	\delta \rho^+ - \delta \rho^- = I M \left( -\sum_{i=1}^\infty f'(\epsilon_i - \epsilon_F) \rho_{ii} \right) .
\end{align}
Integrating the above equation over the unit cell yields
\begin{align}
		M(1- ID_0) = 0 ,
\end{align}
with $D_0 = - \sum_{i=1}^\infty f'(\epsilon_i - \epsilon_F)$ the reference density of states at the Fermi level, which recovers the first-order term of the Stoner equation $M=G(M)$, where a phase transition is only possible at the Stoner criterion, $1 -  ID_0=0$.

\subsection{$\chi_0^{\NAME}$ on supercells}

To evaluate the range of applicability of the proposed model, $\chi_0^\NAME$, we examine its behavior in supercells.

Consider a supercell $\Omega_\mathcal{N}$ with periodic boundary conditions, composed of $\mathcal{N}$ repetitions of the unit cell $\Omega$, in which the external potential $V_\mathrm{ext}$ is $\Omega$-periodic. The Bloch theorem ensures that both the orbital densities $\rho_{ii}^\alpha$ and the local densities of states $D_\loc^\alpha$ are also $\Omega$-periodic. 
Hence, if we consider a potential perturbation $\delta V$ over the supercell, without $\Omega$-periodic component, the scalar products $\braket{\rho^\alpha_{ii}}{\delta V^\alpha}$ and $\braket{D_\loc^\alpha}{\delta V^\alpha}$ vanish, and therefore
\begin{align}
	\chi^\NAME_0 \delta V = 0 .
\end{align}

Put differently, the model $\chi_0^\NAME$ acts exclusively in $\Omega$, and will fail to capture any effect taking place across unit cells.

In particular, this diagonal model will not be able to treat charge sloshing issues, as they are caused by long-range modes that extend beyond the unit cell.
Additionally, it will fail to treat convergence issues caused by small antiferromagnetic modes. Indeed, such modes are characterized by spin densities whose sign alternates between neighboring atoms, and their periodicity exceeds that of $\Omega$.

\subsection{Hybrid preconditioner $P^\PNAME$}

Convergence issues caused by the long-range divergence of the Coulombic interaction happen systematically in metallic systems with large simulation cells. A preconditioner that fails to address these convergence issues is therefore of limited utility.

In order to build a robust preconditioner treating both convergence issues caused by charge sloshing and by small ferromagnetic modes, we propose to combine the model $\chi_0^\NAME$ with the LDOS preconditioner developed in~\cite{herbst_black-box_2020}. The LDOS preconditioner is specifically designed to address charge sloshing problems. Within the collinear spin framework, the non-interacting susceptibility model $\chi_0^\LDOS$ of the LDOS preconditioner is given by the integral kernels
\begin{align}
	\begin{split}
	\left(\chi_0^\LDOS\right)^{\alpha \beta}(\rr, \rr')
	= &-\delta_{\alpha \beta} D_\loc^\alpha(\rr)\delta(\rr-\rr') \\
	&+ \frac{1}{D^+ + D^-} D_\loc^\alpha(\rr) D_\loc^\beta(\rr')
	\end{split}
\end{align}
for $\alpha, \beta \in \{+, -\}$. This model uses the local density of states (LDOS) $D_\loc^\alpha$, which localizes the preconditioner, making it suitable for inhomogeneous systems. The spin channel representation shows that $\chi_0^\LDOS$ is also ``local'' in spin as it uses the corresponding local density of states for each spin channel. This allows the LDOS preconditioner to capture different behaviors in each spin channel. For instance, it is well adapted for half-metals, that are metallic in one spin channel and insulating in the other.

Since the two models $\chi_0^\LDOS$ and $\chi_0^\NAME$ are designed to capture different modes, respectively, stiff long-range modes arising from the Coulombic interaction and soft short-range ferromagnetic modes arising from the exchange and correlation, it is reasonable to use the following hybrid preconditioner:
\begin{align}
	P^\PNAME = 1 - \chi_0^\LDOS K_H - \chi_0^\NAME K_\XC .
\end{align}
This is the final expression of the proposed preconditioner, that we will refer to as the hybrid preconditioner in the rest of the paper. Its performance is tested in Section~\ref{sec:results}.

\subsection{Implementation details and computation cost}\label{sec:implementation}

In SCF iterations, we need to apply the inverse of the preconditioner $P^\PNAME$. This inversion is done iteratively with the GMRES~\cite{saad_gmres_1986} algorithm. 

The $\chi_0^\NAME = \chi_0^{\delta\epsilon} + \chi_0^{\delta\epsilon_F}$ operator is applied with the following formula for $\chi_0^{\delta\epsilon}$:
\begin{align}
	\left( \chi_0^{\delta\epsilon} \delta V \right) ^\alpha = \sum_{i=1}^\infty f'(\epsilon_{i}^\alpha-\epsilon_F) \braket{\rho_{ii}^\alpha}{\delta V^\alpha} \rho_{ii}^\alpha ,
		\label{eq:chi0_diag_apply}
\end{align}
where the quantity $f'(\epsilon_i^\alpha - \epsilon_F)$ is non-negligible only for $\epsilon_i^\alpha$ close to the Fermi level. If $p$ represents the fraction of computed bands sufficiently close to the Fermi level (for which $f'(\epsilon_i^\alpha - \epsilon_F)$ is non-negligible), computing the application of $\chi_0^\NAME$ to a vector has a cost similar to $p$ times the cost of computing the density from the wavefunctions $\psi_i$. 

\begin{figure}
		\centering
		\includegraphics[width=\linewidth]{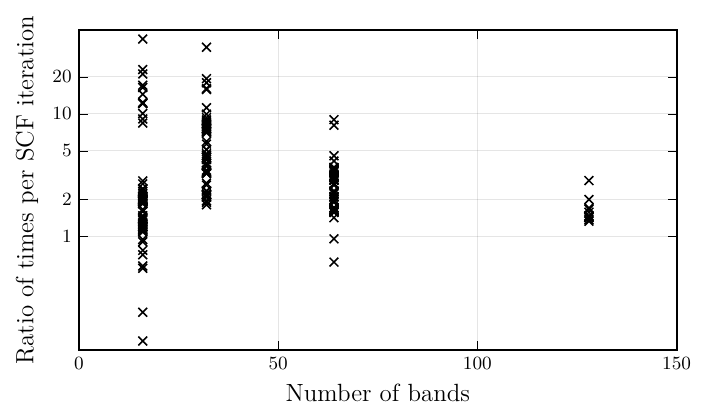}
		\caption{Ratio of CPU times per SCF iteration between the hybrid preconditioner and the Kerker preconditioner for a range of \textsc{Abinit} SCF runs (FFT results not saved).}
    \label{fig:time_ratio}
\end{figure}

In plane-wave KS-DFT codes, the computational cost of one self-consistent iteration is dominated by the diagonalization step, which scales as $\O(N^3)$, where $N$ is the number of electrons. The application of $\chi_0^\NAME$, on the other hand, is dominated by the FFTs and scales as  $\O(N^2 \log(N))$, as computing the orbital densities $\rho_{ii}$ involves transforming $\psi_i$ from the reciprocal space to the direct space. The preconditioner cost becomes negligible for large systems but remains significant for smaller ones. This is illustrated in Figure~\ref{fig:time_ratio}, which shows the ratio of the time per SCF iteration between the hybrid preconditioner and the Kerker preconditioner in \textsc{Abinit} for a range of SCF runs. We observe that as the number of bands increases, the overhead caused by the hybrid preconditioning scheme decreases.

Since $\chi_0^\NAME$ must be applied at each GMRES iteration, there is a tradeoff to find between the computational cost of applying the same FFT several times and the memory cost of saving the result of these FFTs throughout the GMRES iterations. If the FFT results are saved, the computational cost does not significantly increase compared to the use of the LDOS preconditioner alone. If the FFT results are not saved, the preconditioner will have a cost similar to $p \cdot n_\mathrm{GMRES}$ density calculations, and we have found that limiting the number of GMRES iterations $n_\mathrm{GMRES}$ to 30 yields satisfactory results.

This maximum number of GMRES iterations also serves as a safeguard in cases where the hybrid preconditioner is itself singular or ill-conditioned. In practice, we do not expect this to happen, because the diagonal $\chi_0$ model neglects a negative term ($\chi_0^{\delta\psi}$), associated with a negative $K_\XC$ on problematic modes. For this reason, we expect the smallest eigenvalue of the hybrid preconditioner to remain larger than the smallest eigenvalue of $\varepsilon^\dagger$, which is positive at stable solutions of the Kohn-Sham equations. This is verified on the FCC iron system, in which the smallest eigenvalue of $P^\PNAME$ is always larger than that of $\varepsilon^\dagger$ (Figure~\ref{fig:eigmin_precon_wrt_lsf_high_temp}). Still, this heuristic argument  does not guarantee a well-conditioned preconditioner and the maximum number of GMRES iterations remains a necessary safeguard. We have also noticed that, like the LDOS preconditioner but to a greater extent, the hybrid preconditioner is very sensitive to the convergence in k-points. Indeed, the operator $\chi_0$ is more challenging to converge in k-points than other KS-DFT quantities. To overcome this difficulty, we propose to artificially increase the smearing temperature in the preconditioner to ensure its smoothness. This introduces one parameter in our otherwise parameter-free preconditioner. We used a preconditioner smearing temperature of 0.01 Hartree for all our tests. We do a sensitivity analysis to this parameter in Section~\ref{sec:results}.

The hybrid preconditioner has been implemented in both DFTK~\cite{herbst_dftk_2021} and \textsc{Abinit} 10.8~\cite{verstraete_abinit_2025}. In DFTK, the LDOS preconditioner was already available, along with the framework for $\chi_0$-based preconditioners. This framework is compatible with the rest of DFTK's features: in particular, this new preconditioner is compatible with GPU acceleration and can be used within DFPT calculations. In \textsc{Abinit}, we have created a new preconditioning structure, to support $\chi_0$-based preconditioners. The diagonal $\chi_0$ model is implemented as a matrix-free operator described in equation~\eqref{eq:chi0_diag_apply}, with the two options for handling the FFTs (saving or recomputing them) available. The implementation is compatible with most of \textsc{Abinit}'s other features, including the PAW framework, non-collinear magnetism and parallelization over spin, k-points and bands. To invert the preconditioner, we used an existing Fortran90 GMRES solver from~\cite{willis_openpipeflow_2017}.

\section{Numerical performance}\label{sec:results}

In this section, we present the results obtained with the proposed preconditioner $P^\PNAME$ on different magnetic systems. We start by analyzing in detail the behavior of the hybrid preconditioner on FCC iron, whose phase transition we studied in Section~\ref{sec:convergence}. Then we present numerical results on a wider variety of systems and compare the performance of the hybrid preconditioner with the widely used Kerker preconditioner~\cite{kerker_efficient_1981} and the LDOS preconditioner~\cite{herbst_black-box_2020}, of which $P^\PNAME$ is an extension.

\subsection{Preconditioning FCC iron with $P^\PNAME$}

\begin{figure}
	\centering
	\includegraphics[width=\linewidth]{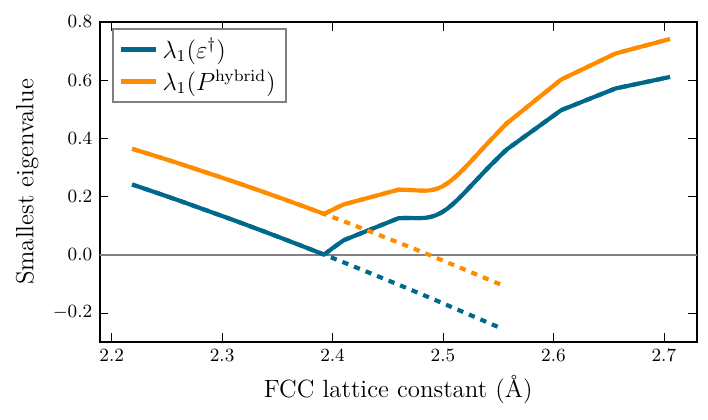}
	\caption{Comparison of the smallest eigenvalue of the dielectric matrix $\varepsilon^\dagger$ and the preconditioner $P^\PNAME$ for the FCC iron system, for various lattice constants.}
    \label{fig:eigmin_precon_wrt_lsf_high_temp}
\end{figure}

For the hybrid preconditioner to efficiently address convergence issues related to small magnetic modes, it must accurately reproduce the small modes of the dielectric matrix. In Figure~\ref{fig:eigmin_precon_wrt_lsf_high_temp}, we represent the smallest eigenvalue ($\lambda_1$) of the preconditioner $P^\PNAME$ and of the adjoint dielectric matrix.
The smallest eigenvalue of $P^\PNAME$ follows closely that of the adjoint dielectric matrix, with a constant discrepancy of approximately 0.1.
We also compute the angle, $\angle(x, y) =  \arccos \left( |\braket{x}{y}| / (\|x\| \|y\|) \right)$, between the two eigenvectors associated with the smallest eigenvalue of $\varepsilon^\dagger$ and $P^\PNAME$. We observe that this angle remains around 10° near the magnetic phase transition.

\begin{figure}
	\centering
	\includegraphics[width=\linewidth]{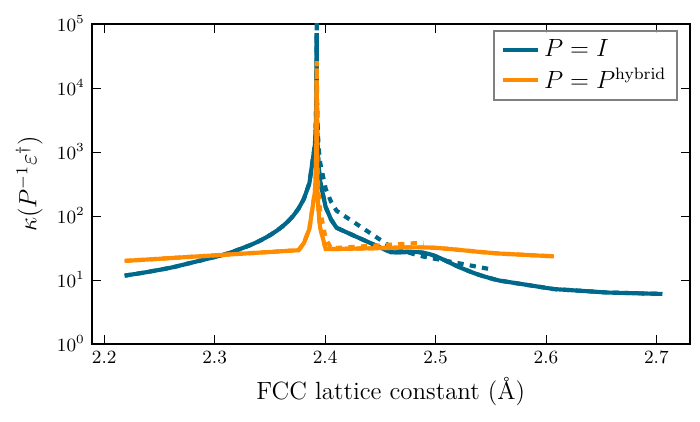}
	\caption{Comparison of the spectral condition number of the adjoint dielectric matrix, not preconditioned ($\kappa_s(\varepsilon^\dagger)$) and preconditioned with the hybrid preconditioner ($\kappa_s((P^\PNAME)^{-1} \varepsilon^\dagger)$) in the FCC iron system with varying lattice constant.}
    \label{fig:cond_precon_wrt_lsf_high_temp}
\end{figure}

This rather good approximation of the smallest eigenmode of the adjoint dielectric matrix by the preconditioner $P^\PNAME$ translates to a reduction of the spectral condition number of the preconditioned adjoint dielectric matrix $(P^\PNAME)^{-1} \varepsilon^\dagger$. Figure~\ref{fig:cond_precon_wrt_lsf_high_temp} shows the spectral condition number of $\varepsilon^\dagger$ and of $\left( P^\PNAME \right)^{-1} \varepsilon^\dagger$. We can see that the hybrid preconditioner significantly reduces the spectral condition number of the preconditioned dielectric matrix near the magnetic transition, where the unpreconditioned dielectric matrix is ill-conditioned. Away from the transition, the behavior of the hybrid preconditioner is governed by the LDOS preconditioner. In this specific example, with no charge sloshing issues, the LDOS preconditioner happens to worsen the condition number of the dielectric matrix.

\begin{figure}
	\centering
	\includegraphics[width=\linewidth]{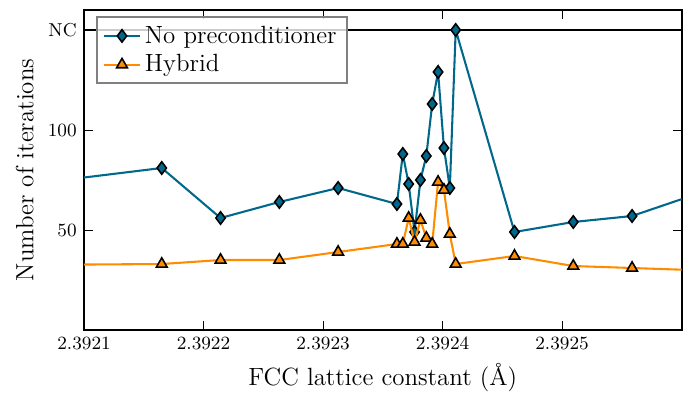}
	\caption{Number of SCF iterations needed to reach a density residual smaller than $10^{-10}$ with respect to the lattice constant for FCC iron. The calculations used Anderson acceleration with a history of length $m=10$ and either no preconditioner or the hybrid preconditioner. NC indicates that the system did not converge after 150 iterations.}
    \label{fig:nite_stoner_sm_wrt_lsf_high_temp}
\end{figure}

\begin{figure}
	\centering
	\includegraphics[width=\linewidth]{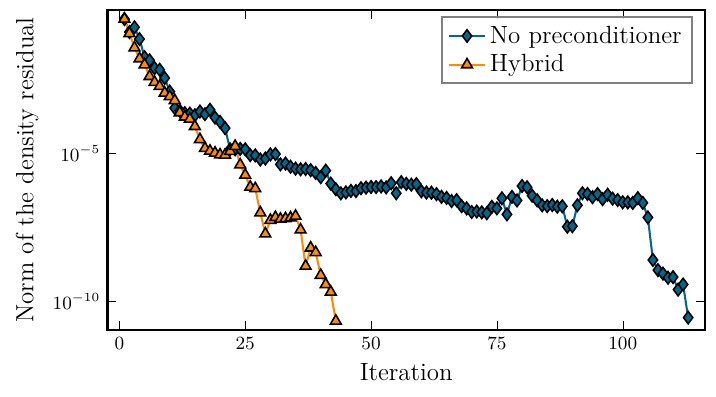}
	\caption{Residual norm of the density throughout the SCF iterations for FCC iron with a lattice constant of 2.3923913 Å, with either no preconditioning or the hybrid preconditioner. The calculations used Anderson acceleration with a history of length $m=10$.}
    \label{fig:convergence_sm_Fe_high_temp_transition}
\end{figure}

This reduced spectral condition number around the transition translates into a significantly smaller number of SCF iterations at the transition, as illustrated in Figure~\ref{fig:nite_stoner_sm_wrt_lsf_high_temp}. Figure~\ref{fig:convergence_sm_Fe_high_temp_transition} shows an example of a convergence plot at the transition, on which we see that the hybrid preconditioner eliminates the plateau. Thus, the hybrid preconditioner seems appropriate to address convergence issues in FCC iron.

\subsection{$P^\PNAME$ on a wider range of magnetic systems}

We now present numerical results on a wider variety of magnetic systems. The DFT simulations presented here were performed with the \textsc{Abinit} software~\cite{verstraete_abinit_2025}, with a plane-wave basis set and norm-conserving pseudopotentials~\cite{van_setten_pseudodojo_2018}. We used the same cutoff energy of 50 Hartree for all systems and a Monkhorst-Pack k-point grid. The exchange-correlation functional used is the GGA PBE functional~\cite{perdew_generalized_1996}, combined with a Fermi-Dirac occupation scheme. For all systems, convergence with respect to the cutoff energy and k-point grid ensures a maximum error of $10^{-3}$ Ha in the total energy and a maximum relative error of $2\%$ in the magnetization. 

We consider a range of ferromagnetic systems to compare the efficiency of the Kerker preconditioner, the LDOS preconditioner, and the hybrid preconditioner. We chose systems with a small total magnetization per unit cell, and we adjusted the lattice constants of these systems to bring them closer to a magnetic phase transition, where small magnetic modes are expected to appear in the dielectric matrix. All SCF procedures were accelerated using Pulay mixing with a history size of 10 and a damping parameter of 0.8. The Kohn-Sham eigenproblem was solved iteratively with a degree-16 Chebyshev polynomial filter.

\begin{figure}
	(1) \ce{Co}\\
	\includegraphics[width=\linewidth]{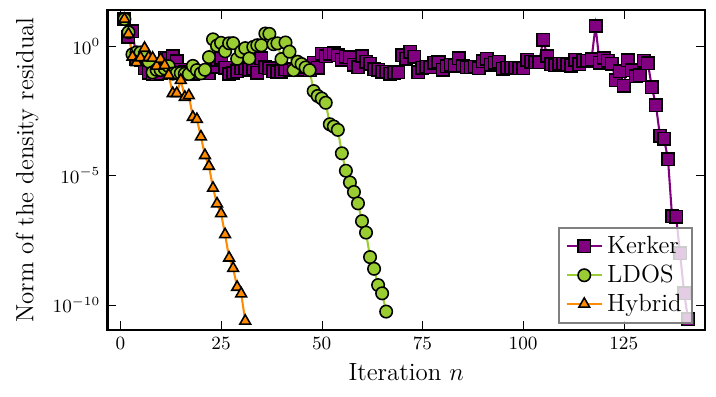}
	(2) \ce{Ni}\\
	\includegraphics[width=\linewidth]{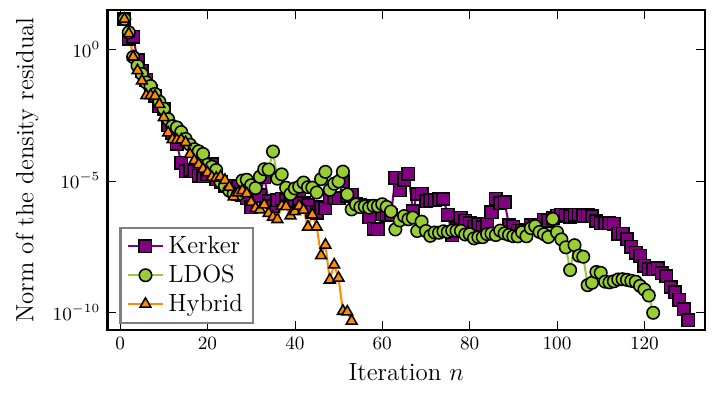}
	(3) \ce{ZrZn2}\\
	\includegraphics[width=\linewidth]{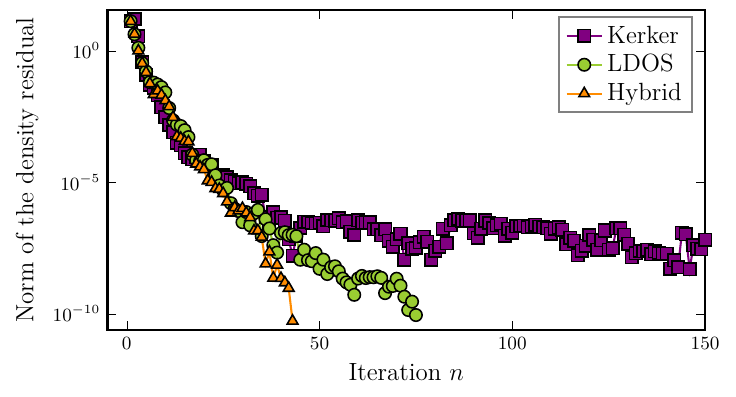}
	\caption{Convergence plots for the systems FCC cobalt, FCC nickel and \ce{ZrZn2}, with the Kerker preconditioner, the LDOS preconditioner, and the hybrid preconditioner. The convergence curves may vary across different computer architectures, but the overall trend (whether a plateau is present or not) remains robust.}
    \label{fig:conv_abinit}
\end{figure}

When the system parameters are intentionally adjusted to place the system near a phase transition, we observe significant accelerations with the hybrid preconditioner. Figure~\ref{fig:conv_abinit} shows the convergence plots for three systems in which $P^\PNAME$ significantly accelerates convergence, FCC 
\ce{Co}\footnote{\ce{Co} [mp-102] with a $49\times49\times49$ k-point grid, a smearing temperature of 0.001 Ha and the lattice constants scaled by a factor 0.91 (convergence issues from 0.908 to 0.915).},
FCC \ce{Ni}\footnote{\ce{Ni} [mp-23] with a $49\times49\times49$ k-point grid, a smearing temperature of 0.005 Ha and the lattice constants scaled by a factor 0.830104 (convergence issues from 0.830096 to 0.830106).}
and \ce{ZrZn2}\footnote{\ce{ZrZn2} [mp-1401] with a $25\times25\times25$ k-point grid, a smearing temperature of 0.005 Ha and the lattice constants scaled by a factor 0.9332 (convergence issues from 0.9312 to 0.9336).}, 
whose structures are taken from the Materials Project~\cite{jain_commentary_2013}. Table~\ref{tab:times} also shows the effective times (total and per SCF iteration) of these calculations. For these systems, we observe a plateau in the convergence curves obtained with the Kerker and LDOS preconditioners, characteristic of the presence of small eigenvalues in the dielectric matrix. The hybrid preconditioner either eliminates this plateau or significantly reduces its size. Still, we observe that $P^\PNAME$ does not solve all convergence issues caused by ferromagnetic phase transitions. For instance, when the FCC cobalt system is brought even closer to its transition, it does not converge with any of the preconditioners, including the hybrid scheme. Moreover, for \ce{ZrZn2}, the reduction in the number of iterations is not large enough to compensate for the additional cost per iteration of the hybrid preconditioner.

\begin{table}
		\centering
		\begin{tabular}{llccc}
			\toprule
			& & \ce{Co} & \ce{Ni} & \ce{ZrZn2} \\
			\midrule
			\multirow{3}{*}{\shortstack[l]{Time per SCF\\iteration (h)}}
			& Kerker & 0.623 & 0.509 & 6.01 \\
			& LDOS   & 0.632 & 0.508 & 6.15 \\
			& Hybrid & 0.758 & 0.615 & 27.4 \\
			\midrule
			\multirow{3}{*}{Total time (h)}
			& Kerker & 87.8 & 66.2 & NC \\
			& LDOS   & 41.7 & 62.0 & 462 \\
			& Hybrid & 25.0 & 32.6 & 1,180 \\
			\bottomrule
		\end{tabular}
		\caption{Total time and time per SCF iteration (total CPU-hours) for the three systems presented in Figure~\ref{fig:conv_abinit} (FFT results not saved).}
    \label{tab:times}
\end{table}

\begin{figure}
	\includegraphics[width=\linewidth]{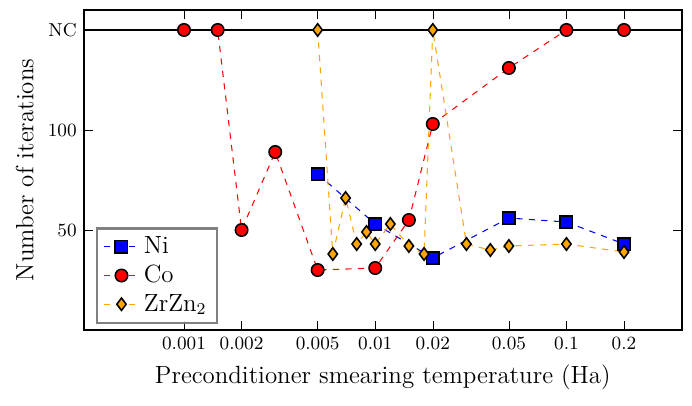}
	\caption{Number of SCF iterations needed to reach a density residual below $10^{-10}$ with respect to the hybrid preconditioner smearing temperature, for FCC cobalt, FCC nickel, and \ce{ZrZn2}. NC indicates that the calculation did not converge after 150 iterations.}
	\label{fig:impact_ptsmear}
\end{figure}

The results presented in Figure~\ref{fig:conv_abinit} use a fixed preconditioner smearing temperature of 0.01~Ha for the hybrid preconditioner. We have tested the sensitivity of the hybrid preconditioner to its smearing temperature on these same systems. Figure~\ref{fig:impact_ptsmear} shows the number of SCF iterations required for convergence as a function of the preconditioner smearing temperature. The sensitivity varies from one system to another. For nickel, the smearing temperature has little impact on convergence. For cobalt, the hybrid preconditioner becomes ineffective if the smearing temperature is set too high, with the range of acceptable values spanning roughly one order of magnitude. For \ce{ZrZn2}, most smearing temperatures work well, but the hybrid preconditioner shows some instability, as a few specific smearing temperatures cause it to fail.

In systems simulated away from their magnetic phase transition, which present no convergence issues related to magnetism, the hybrid preconditioner generally has no significant effect and does not worsen the convergence of the SCF iterations. 
However, the hybrid preconditioner is more sensitive to the convergence in k-points. In our experiments, we have found a few systems, with insufficient k-point sampling, for which the use of $P^\PNAME$ significantly worsens the convergence, even with the increased smearing temperature in the preconditioner mentioned in Section~\ref{sec:implementation}.

The newly proposed hybrid preconditioning scheme thus appears promising to address convergence issues related to magnetism. We successfully accelerated calculations for the three typical ferromagnetic elements, iron, cobalt, and nickel, as well as an alloy, \ce{ZrZn2}. Still, the applicability of this preconditioning scheme remains limited to ferromagnetic phase transitions.

\section{Conclusion}

We have investigated convergence issues in magnetic systems and demonstrated that, although these convergence issues are localized around phase transitions, they can be significant. Inspired by the Stoner model, we derived a model that targets problematic ferromagnetic modes, combining it with the LDOS preconditioner to address both magnetic and charge sloshing issues. This new preconditioner is available in both DFTK and \textsc{Abinit}~10.8.

Our numerical results show that the proposed hybrid preconditioner performs well within its narrow range of applicability, namely ferromagnetic systems close to a second-order phase transition. Outside this range, it generally does not degrade convergence, but it does increase the cost per SCF iteration. For these reasons, we do not recommend using it as a default choice. Instead, we suggest using it for ferromagnetic systems whose SCF convergence exhibits a plateau. In our experience, a preconditioner smearing temperature of 0.01~Ha works well, and we recommend adjusting this value if convergence is not improved with the default setting. Finally, we recall that the hybrid preconditioner is not expected to be effective for antiferromagnetic systems, and that it may behave unpredictably when the k-point sampling is insufficient.

Overall, this study provides a deeper understanding of convergence issues in spin-KS-DFT and offers a practical solution for improving the robustness of SCF iterations in challenging systems.

Another issue commonly encountered in magnetic systems is the possible existence of multiple stable solutions to the Kohn-Sham equations. Although preconditioning can influence which solution the SCF iterations converge to, the problem of identifying the physically relevant solution is largely independent of accelerating convergence toward a given solution. A natural extension to this work would be to explore the solution landscapes in magnetic systems.

\bibliographystyle{elsarticle-num}
\bibliography{biblio_article_magnetisme}

\end{document}